\documentclass[a4paper,11pt]{article}

\usepackage{amsmath,amssymb,bbm}
\usepackage{ascmac}
\usepackage{cancel}
\usepackage{graphicx}
\usepackage[bookmarks=true,bookmarksnumbered=true,bookmarkstype=toc]{hyperref}
\hypersetup{
pdfauthor={Tetsuji Kimura},
colorlinks={true},
linkcolor={blue},
urlcolor={blue},
filecolor={blue},
citecolor={blue}
}
\usepackage{ulem}

\makeatletter

 \@addtoreset{equation}{section}
\makeatother

\newcounter{Enumerate}

\DeclareFontFamily{U}{rsf}{}
\DeclareFontShape{U}{rsf}{m}{n}{
  <5> <6> rsfs5 <7> <8> <9> rsfs7 <10-> rsfs10}{}
\DeclareMathAlphabet\Scr{U}{rsf}{m}{n}

\usepackage[mathscr]{eucal}

\newcommand{\half}{\frac{1}{2}}

\newcommand{\ls}{\ \ \ \ \ }
\newcommand{\wt}{\widetilde}
\newcommand{\wh}{\widehat}

\newcommand{\bsubeq}{\begin{subequations}}
\newcommand{\esubeq}{\end{subequations}}

\newcommand{\noi}{\noindent}

\newcommand{\I}{{\rm i}}
\newcommand{\N}{\mathcal{N}}
\newcommand{\T}{{\rm T}}

\renewcommand{\d}{{\rm d}}
\newcommand{\e}{{\rm e}}

\newcommand{\slb}{\scalebox}

\begin{document}
\allowdisplaybreaks{

\thispagestyle{empty}


\begin{flushright}
TIT/HEP-640 \\
\vphantom{last update:} 
\end{flushright}

\vspace{40mm}

\noi
\slb{2.5}{Defect $(p,q)$ Five-branes}

\vspace{18mm}

\slb{1.2}{Tetsuji {\sc Kimura}}

\slb{1}{\renewcommand{\arraystretch}{1.2}
\begin{tabular}{rl}
& 
{\sl 
Department of Physics,
Tokyo Institute of Technology} 
\\
& {\sl 
Tokyo 152-8551, JAPAN
}
\\
& {\tt tetsuji \_at\_ th.phys.titech.ac.jp}
\end{tabular}
}

\vspace{35mm}


\slb{1.1}{\sc Abstract}
\begin{center}
\slb{1}{
\begin{minipage}{.9\textwidth}
We study 
a local description of composite five-branes of codimension two.
The formulation is constructed by virtue of 
$SL(2,{\mathbb Z}) \times SL(2,{\mathbb Z)}$ monodromy associated with two-torus.
Applying conjugate monodromy transformations to the complex structures of the two-torus, 
we obtain a field configuration of a defect $(p,q)$ five-brane.
This is a composite state of $p$ defect NS5-branes and $q$ exotic $5^2_2$-branes. 
We also obtain a new example of
hyper-K\"{a}hler geometry.
This is an ALG space, a generalization of an ALF space which asymptotically has a tri-holomorphic two-torus action.
This geometry appears in the conjugate configuration of 
a single defect KK5-brane.
\end{minipage}
}
\end{center}


\newpage
\section{Introduction}
\label{sect:introduction}

A Neveu-Schwarz five-brane, called an NS5-brane for short, plays a significant role in string theory.
This is a soliton coupled to 
B-field magnetically in ten-dimensional spacetime, 
whereas a fundamental string is coupled to the B-field electrically \cite{Strominger:1990et, Callan:1991dj}.
A setup of two parallel NS5-branes with various D-branes attached with them is quite an important configuration to explore dualities among supersymmetric gauge theories 
\cite{Hanany:1996ie, Giveon:1998sr}. 
An NS5-brane is uplifted to an M5-brane in M-theory, which plays a central role in studying non-perturbative features of gauge theories in lower dimensions \cite{Gaiotto:2009we}.
Applying T-duality to an NS5-brane along one transverse direction, 
a Kaluza-Klein monopole \cite{Sorkin:1983ns}, 
or referred to as a KK5-brane, emerges.
If one performs T-duality to the KK5-brane 
along another transverse direction,
one finds an exotic $5^2_2$-brane \cite{deBoer:2010ud}.
This is a strange object whose background geometry is no longer single-valued.
Furthermore, this strange object does also contribute to quantum aspects of spacetime \cite{deBoer:2012ma}.

NS5-branes and KK5-branes have been investigated from various viewpoints \cite{Gregory:1997te}.
In particular, in order to analyze quantum stringy corrections to five-branes,
the worldsheet approach to five-branes \cite{Callan:1991dj} has been developed in terms of two-dimensional supersymmetric gauge theory, called the gauged linear sigma model (GLSM) \cite{Hori:2002cd, Tong:2002rq, Harvey:2005ab, Okuyama:2005gx}.
In the case of an exotic $5^2_2$-brane, the situation is different.
The background geometry is written by a multi-valued function 
because an exotic $5^2_2$-brane is codimension two.
These days, branes of codimension two are referred to as {\it defect} branes \cite{Bergshoeff:2011se}.
The exotic $5^2_2$-brane is 
a typical example of defect five-branes.
Indeed, it was difficult to construct both the worldsheet theory and 
the worldvolume theory for 
an exotic $5^2_2$-brane.
However, there was a breakthrough in this topic.
The GLSM for an exotic $5^2_2$-brane was successfully obtained in \cite{Kimura:2013fda}.
This formulation enables us to study 
quantum aspects of an exotic $5^2_2$-brane \cite{Kimura:2013zva} in the same way as NS5-branes and KK5-branes
\cite{Tong:2002rq, Harvey:2005ab, Okuyama:2005gx}.
The worldvolume theory for an exotic $5^2_2$-brane was also constructed \cite{Chatzistavrakidis:2013jqa, Kimura:2014upa} by following the work \cite{Bergshoeff:1997gy}.

In the analyses of five-branes, 
people often encounter many of their configurations.
A typical example is a defect $(p,q)$ five-brane.
This is a composite state of $p$ defect NS5-branes and $q$ exotic $5^2_2$-branes \cite{deBoer:2010ud, deBoer:2012ma}.
This is one of the most significant situations to formulate globally well-defined description of defect five-branes.
This resembles a $(p,q)$ seven-brane in type IIB theory \cite{Greene:1989ya, DeWolfe:1998eu, DeWolfe:1998pr, Bergshoeff:2006jj}.
It has been argued as a globally well-defined description of a defect $(p,q)$ five-brane in terms of the modular $J$ function \cite{Hellerman:2002ax, Kikuchi:2012za}. 
This is related to a globally well-defined description of a $(p,q)$ seven-brane via string dualities.
It should be important to find a {\it direct} derivation of 
a globally well-defined description of a defect $(p,q)$ five-brane without the aid of seven-branes.

It is quite an important task to construct 
a globally well-defined description of a defect $(p,q)$ five-brane.
In order to complete this, 
we study its ``local'' description {\it as the first step}.
In this paper, we exhaustively utilize monodromy structures of 
a defect five-brane.
Applying aspects of the monodromy to its background fields,
we obtain an explicit form of a defect $(p,q)$ five-brane.
Even though the formulation tells us only the {\it local} structure of the five-branes, it would be a big step to find the globally well-defined form.
%
In addition,
we find a new example of hyper-K\"{a}hler geometry, as a bonus.
This is called an ALG space \cite{Cherkis:2000cj, Cherkis:2001gm}, 
a generalization of an ALF space which asymptotically has a tri-holomorphic two-torus action.
This is the conjugate geometry of a defect KK5-brane.

The structure of this paper is as follows.
In section \ref{sect:review}, 
we review standard five-branes and defect five-branes.
First, we exhibit their local descriptions.
Next, we discuss $O(2,2;{\mathbb Z})$ monodromy
of the defect five-branes
and mention a nongeometric feature.
In section \ref{sect:SL2SL2-monodromy},
we further study the monodromy
of the defect five-branes by virtue of the equivalence $O(2,2;{\mathbb Z}) = SL(2,{\mathbb Z}) \times SL(2,{\mathbb Z})$.
We introduce two complex structures associated with two $SL(2,{\mathbb Z})$.
They are the key ingredients to analyze composites of defect five-branes.
In section \ref{sect:ConjugateConfig},
we investigate conjugate monodromies and construct their corresponding configurations.
In particular, we obtain a local description of a defect $(p,q)$ five-brane.
This is a composite of $p$ defect NS5-branes and $q$ exotic $5^2_2$-branes.
We also obtain the conjugate configuration of a defect KK5-brane.
This provides a new example of
hyper-K\"{a}hler geometry as an ALG space.
Section \ref{sect:summary} is devoted to summary and discussions.
In appendix \ref{app:Buscher},
we prepare the T-duality transformation rules applied to the field configurations and monodromy matrices.
In appendix \ref{app:AK5},
we discuss another defect KK5-brane which is different from the reduction of the standard KK5-brane, and analyze its conjugate configuration.

\section{A review of defect five-branes}
\label{sect:review}

\subsection{Standard five-branes}
\label{sect:SB}

In this subsection we briefly mention explicit descriptions of an H-monopole and a KK-monopole \cite{Gregory:1997te}.
An H-monopole is an NS5-brane smeared
along one of the transverse direction,
while a KK-monopole is 
a five-brane generated by T-duality along the smeared direction of the H-monopole.
These two objects have been well investigated in the framework of 
GLSM \cite{Tong:2002rq, Harvey:2005ab, Okuyama:2005gx},
and doubled formalism \cite{Jensen:2011jna, Berman:2014jsa}
(see also \cite{deBoer:2012ma, Andriot:2014uda}).


We begin with the H-monopole.
In ten-dimensional spacetime, 
we describe the background metric $G_{MN}$, the B-field $B_{MN}$ and the dilaton $\phi$ as
\bsubeq \label{H-system}
\begin{gather}
\d s^2 \ = \ 
\d s_{012345}^2
+ H \big[ (\d x^6)^2 + (\d x^7)^2 + (\d x^8)^2 + (\d x^9)^2 \big]
\, , \\
B_{i9} \ = \ V_i
\, , \ls
\e^{2 \phi} \ = \ H
\, , \\
H \ = \ 
1 + \frac{\ell_0}{\sqrt{2} \, |\vec{x}|}
\, , \ls
\ell_0 \ = \ \frac{\alpha'}{R_9}
\, , \\
\nabla_i H \ = \ (\nabla \times \vec{V})_i
\, , \ls
\vec{V} \cdot \d \vec{x}
\ = \ 
\frac{\ell_0}{\sqrt{2}} 
\frac{- x^6 \d x^8 + x^8 \d x^6}{|\vec{x}| (|\vec{x}| + x^7)} 
\, . \label{monopole-eq}
\end{gather}
\esubeq
Here $\alpha'$ is the Regge parameter in string theory.
The NS5-brane is expanded in the 012345-directions whose spacetime metric is flat, while the transverse space of the 6789-directions is ${\mathbb R}^3 \times S^1$.
The vector $\vec{x}$ lives in the transverse 678-directions ${\mathbb R}^3$.
This five-brane is smeared along the transverse 9-th compact direction whose radius is $R_9$.
This configuration is governed by a harmonic function $H$.
The B-field is given by a function $V_i$ which is subject to the monopole equation (\ref{monopole-eq}), where the index $i$ represents the spatial directions $i = 6,7,8$.
We also evaluate the mass of the single H-monopole (see, for instance, \cite{Kimura:2014upa}):
\begin{align}
M_{\text{H-monopole}} \ &= \ 
\frac{1}{g_{\text{st}}^2 \ell_{\text{st}}^6}
\, , \label{massH}
\end{align}
where $g_{\text{st}}$ and $\ell_{\text{st}} = \sqrt{\alpha'}$ are the string coupling constant and the string length respectively.


Next, we consider the KK-monopole, or referred to as the KK5-brane.
This is obtained via the T-duality transformation (see appendix \ref{app:Buscher}) along the smeared direction of the H-monopole (\ref{H-system}),
\bsubeq \label{KK5-system}
\begin{gather}
\d s^2 \ = \ 
\d s_{012345}^2
+ H \big[ (\d x^6)^2 + (\d x^7)^2 + (\d x^8)^2 \big] 
+ \frac{1}{H} \big[ \d y^9 - \vec{V} \cdot \d \vec{x} \big]^2
\, , \\
B_{MN} \ = \ 0
\, , \ls
\e^{2 \phi} \ = \ 1
\, .
\end{gather}
\esubeq
Due to the T-duality transformation, the B-field in the H-monopole (\ref{H-system}) is involved into the off-diagonal part of the metric as the KK-vector $\vec{V}$.
We also see that the dilaton becomes trivial.
The transverse space of the 6789-directions becomes the Taub-NUT space, a non-compact hyper-K\"{a}hler geometry.
In order to emphasize the T-duality transformation along the 9-th direction,
we refer to this coordinate as $y^9$ whose radius is $\wt{R}_9$.
Under the T-duality transformation along the $i$-th direction, the radius $R_i$ and the coupling constant $g_{\text{st}}$ are changed as
\begin{align}
R_i \ \to \ \frac{\ell_{\text{st}}^2}{R_i} \ = \ \wt{R}_i
\, , \ls
g_{\text{st}} \ \to \ \frac{\ell_{\text{st}}}{R_i} \, g_{\text{st}}
\, . \label{T-dual-gl}
\end{align}
We should notice that the radius $R_9$ is now that of the {\it dual} coordinate $x^9$. 
The function $H^{-1}$ in front of $(\d y^9)^2$ in (\ref{KK5-system}) asymptotically approaches a dimensionless value $(\wt{R}_9/\ell_{\text{st}})^2$ in the large $|\vec{x}|$ limit.
We obtain the mass of the single KK5-brane via the transformation (\ref{T-dual-gl}),
\begin{align}
M_{\text{KK5}} \ &= \ 
\frac{(R_9)^2}{g_{\text{st}}^2 \ell_{\text{st}}^8}
\ = \ 
\frac{1}{g_{\text{st}}^2 \ell_{\text{st}}^4 (\wt{R}_9)^2}
\, . \label{massKK5}
\end{align}

\subsection{Defect five-branes}
\label{sect:DB}

In the previous subsection we mentioned two standard five-branes of codimension three.
It is interesting to consider five-branes of codimension two, i.e., the {\it defect} five-branes \cite{Bergshoeff:2011se}.
We can easily find defect five-branes from H-monopoles
and KK5-branes
if one of the transverse directions is further smeared\footnote{The smearing procedure can be seen in \cite{deBoer:2010ud, Kikuchi:2012za, Kimura:2013fda} and so forth.}.
One of the most interesting defect five-branes is an exotic $5^2_2$-brane.
This has been investigated in the various viewpoints \cite{deBoer:2010ud, Kikuchi:2012za, deBoer:2012ma, Kimura:2013fda}.


We first discuss a defect NS5-brane smeared along the 8-th direction of the H-monopole (\ref{H-system}).
The configuration is given as 
\bsubeq \label{dNS5-system}
\begin{gather}
\d s^2 \ = \ 
\d s_{012345}^2
+ H_{\ell} \, \big[ (\d \varrho)^2 + \varrho^2 (\d \vartheta)^2 \big]
+ H_{\ell} \big[ (\d x^8)^2 + (\d x^9)^2 \big]
\, , \\
B_{89} \ = \ V_{\ell}
\, , \ls
\e^{2 \phi} \ = \ H_{\ell}
\, , \\
\begin{alignat}{3}
H_{\ell} \ &= \ 
h + \ell \log \frac{\mu}{\varrho}
\, , &\ls
V_{\ell} \ &= \ \ell \vartheta
\, , &\ls
K_{\ell} \ &= \ (H_{\ell})^2 + (V_{\ell})^2
\, , \\
x^6 \ &= \ \varrho \cos \vartheta
\, , &\ls
x^7 \ &= \ \varrho \sin \vartheta
\, , &\ls
\ell \ &= \ 
\frac{\ell_0}{2 \pi R_8}
\, .
\end{alignat}
\end{gather}
\esubeq
Here $R_8$ is the radius of the compact circle along the smeared 8-th direction.
Now the space of the 89-directions becomes a two-torus $T^{89}$.
We notice that the harmonic function $H$ is reduced to a logarithmic function.
Here $\mu$ is the renormalization scale and $h$ is the bare quantity which 
diverges if we go infinitely away from the five-brane.
In this sense the representation (\ref{dNS5-system}) is valid only close to the defect five-brane.
We note that the mass of the single defect NS5-brane is the same as that of the single H-monopole (\ref{massH}):
\begin{align}
M_{\text{NS}} \ &= \ 
\frac{1}{g_{\text{st}}^2 \ell_{\text{st}}^6}
\, . \label{massdNS5}
\end{align}


There exist two isometries along the 8-th and 9-th directions of the defect NS5-brane (\ref{dNS5-system}).
Taking the T-duality transformation along the 9-th direction $x^9$ to $y^9$,
we obtain a defect KK5-brane,
\bsubeq \label{dKK5-system}
\begin{gather}
\d s^2 \ = \ 
\d s_{012345}^2
+ H_{\ell} \, \big[ (\d \varrho)^2 + \varrho^2 (\d \vartheta)^2 \big]
+ H_{\ell} \, (\d x^8)^2 
+ \frac{1}{H_{\ell}} \big[ \d y^9 - V_{\ell} \, \d x^8 \big]^2 
\, , \\
B_{MN} \ = \ 0
\, , \ls 
\e^{2 \phi} \ = \ 1
\, . 
\end{gather}
\esubeq
This is also found if the KK5-brane of codimension three (\ref{KK5-system}) is smeared along the 8-th direction.
The space of the 89-direction is also a two-torus $T^{89}$.
Here the B-field and the dilaton are again trivial.
The mass of the defect KK5-brane is also the same as that of the single KK5-brane (\ref{massKK5}):
\begin{align}
M_{\text{KK}} \ &= \ 
\frac{(R_9)^2}{g_{\text{st}}^2 \ell_{\text{st}}^8}
\ = \ 
\frac{1}{g_{\text{st}}^2 \ell_{\text{st}}^4 (\wt{R}_9)^2}
\, . \label{massdKK5}
\end{align}
Here $\wt{R}_9$ is the radius of the physical coordinate $y^9$ in the configuration (\ref{dKK5-system}), while $R_9$ is now the radius of the dual coordinate $x^9$.

If we take the T-duality transformation along the 8-th direction instead of the 9-th direction of the defect NS5-brane (\ref{dNS5-system}),
we also find the configuration of another defect KK5-brane of different type.
This will be discussed in appendix \ref{app:AK5}.


Performing the T-duality transformation along the 8-th direction $x^8$ of the defect KK5-brane (\ref{dKK5-system}),
we obtain the configuration of an exotic $5^2_2$-brane \cite{deBoer:2010ud, Kikuchi:2012za, deBoer:2012ma, Kimura:2013fda},
\bsubeq \label{522-system}
\begin{gather}
\d s^2 \ = \ 
\d s_{012345}^2
+ H_{\ell} \, \big[ (\d \varrho)^2 + \varrho^2 (\d \vartheta)^2 \big] 
+ \frac{H_{\ell}}{K_{\ell}} \big[ (\d y^8)^2 + (\d y^9)^2 \big]
\, , \\
B_{89} \ = \ - \frac{V_{\ell}}{K_{\ell}}
\, , \ls
\e^{2 \phi} \ = \ \frac{H_{\ell}}{K_{\ell}}
\, . 
\end{gather}
\esubeq
The space of the 89-directions is again a two-torus $T^{89}$.
Here the B-field and the dilaton are non-trivial as in the configuration of the defect NS5-brane (\ref{dNS5-system}).
However, their features are quite different from the ones in (\ref{dNS5-system}).
Indeed, not only the spacetime metric, but also the B-field and the dilaton are no longer {\it single-valued}.
It is impossible to remove such features by the coordinate transformations or by the B-field gauge transformation.
This is the reason why this configuration is called the ``exotic'' five-brane.
In the next subsection we capture the exotic structure by virtue of monodromy.
Here we also evaluate the mass of the single exotic $5^2_2$-brane obtained from that of the defect KK5-brane (\ref{massdKK5}) via the transformation rule (\ref{T-dual-gl}):
\begin{align}
M_{\text{E}} \ &= \ 
\frac{(R_8 R_9)^2}{g_{\text{st}}^2 \ell_{\text{st}}^{10}}
\ = \ 
\frac{1}{g_{\text{st}}^2 \ell_{\text{st}}^{2} (\wt{R}_8 \wt{R}_9)^2}
\, , \label{mass522}
\end{align}
where $\wt{R}_{8,9} \equiv \ell_{\text{st}}^2/R_{8,9}$ are the radii of the physical coordinates $y^{8,9}$, while $R_{8,9}$ are now radii of the dual coordinates $x^{8,9}$.
The function $H_{\ell}/K_{\ell}$ in front of $(\d y^8)^2 + (\d y^9)^2$ in (\ref{522-system}) asymptotically approaches a dimensionless value $(\wt{R}_8/\ell_{\text{st}})^2 = (\wt{R}_9/\ell_{\text{st}})^2$ in the appropriately large $\varrho$ region.

\subsection{$O(2,2; {\mathbb Z})$ monodromy}
\label{sect:O22-monodromy}


When we go around a defect five-brane along the angular coordinate $\vartheta$ in the 67-plane, 
we can capture monodromy generated by the two-torus $T^{89}$.
The analysis of monodromy is important to investigate the exotic structure of defect five-branes.
Now we package the 89-directions of the metric and 
the B-field in a $4 \times 4$ matrix $\mathscr{M}$ 
\cite{Maharana:1992my}\footnote{The matrix $\mathscr{M}$ is called the moduli matrix. These days it is also referred to as the generalized metric in the framework of generalized geometry and double field theory.},
\bsubeq \label{MM-MM}
\begin{align}
\mathscr{M} (\varrho, \vartheta) \ &\equiv \ 
\left(
\begin{array}{cc}
G_{mn} - B_{mp} \, G^{pq} \, B_{qn} & B_{mp} G^{pn} \\
- G^{mp} B_{pn} & G^{mn}
\end{array}
\right)
\, , \ls
m,n,\ldots = 8, 9
\, .
\end{align}
The matrix $\mathscr{M}$ is restricted to the coset space $O(2,2)/[O(2) \times O(2)]$ \cite{Hull:2004in}.
The numerator $O(2,2)$ is related to the T-duality symmetry $O(2,2;{\mathbb Z})$ on the two-torus $T^{89}$, 
while the denominator $O(2) \times O(2)$ describes the local symmetry related to the coordinate transformations and the B-field gauge transformation.
When we go around a five-brane along the coordinate $\vartheta$ from $0$ to $2 \pi$, the matrix $\mathscr{M}$ is transformed as
\begin{align}
\mathscr{M}(\varrho, 2 \pi)
\ &= \ 
\Omega^{\T} \, \mathscr{M} (\varrho, 0) \, \Omega
\, .
\end{align}
\esubeq
The transformation matrix $\Omega$ indicates the monodromy of the system.
This monodromy takes valued in $O(2,2; {\mathbb Z})$.
We discuss the monodromy matrix more in detail 
\cite{Hull:2004in, Lawrence:2006ma, Hull:2006qs, Dall'Agata:2007sr, Andriot:2014uda}.
The matrix $\Omega$ is described as \cite{Dall'Agata:2007sr}
\begin{align}
\Omega \ &= \ 
\left(
\begin{array}{cc}
A & \beta 
\\
\Theta & D
\end{array}
\right)
\, , 
\end{align}
where $A$, $D$, $\Theta$ and $\beta$ are $2 \times 2$ block matrices.
The blocks $A$ and $D$ govern the coordinate transformations, 
while $\Theta$ gives rise to the B-field gauge transformation.
If the block $\beta$ exists non-trivially, 
the T-duality is involved into the geometrical structure.
A configuration involving $\beta$ in the monodromy matrix
is called a T-fold \cite{Hull:2004in}.
Such a space is locally geometric but globally nongeometric.

Now we explicitly describe
the matrices $\mathscr{M}$ and $\Omega$ of the defect five-branes.
The defect NS5-brane (\ref{dNS5-system}) has the following matrices,
\begin{align}
\mathscr{M}^{\text{NS}} (\varrho, \vartheta)
\ &= \ 
\frac{1}{H_{\ell}}
\left(
\begin{array}{cccc}
K_{\ell} & 0 & 0 & V_{\ell} \\
0 & K_{\ell} & - V_{\ell} & 0 \\
0 & - V_{\ell} & 1 & 0 \\
V_{\ell} & 0 & 0 & 1
\end{array}
\right)
\, , \ls 
\Omega^{\text{NS}}
\ = \ 
\left(
\begin{array}{cccc}
1 & 0 & 0 & 0 
\\
0 & 1 & 0 & 0 
\\
0 & -2 \pi \ell & 1 & 0
\\
2 \pi \ell & 0 & 0 & 1
\end{array}
\right)
\, . 
\label{SO22-MM-dNS5}
\end{align}
We note that the monodromy matrix $\Omega^{\text{NS}}$ contains the $\Theta$ part, while it does not contain the $\beta$ part.
This is consistent with the configuration (\ref{dNS5-system}), 
where the $2 \pi$ shift of the coordinate $\vartheta$ is removed by the B-field gauge transformation.
When we move around the defect NS5-brane $\vartheta = 0 \to 2 \pi n$ with $n \in {\mathbb Z}$, the monodromy matrix 
is given by $(\Omega^{\text{NS}})^n$. 
This is equal to $\Omega^{\text{NS}}$ whose components $\pm 2 \pi \ell$ are replaced to $\pm 2 \pi \ell n$.
In the same way, we study the matrices of the defect KK5-brane (\ref{dKK5-system}),
\begin{align}
\mathscr{M}^{\text{KK}} (\varrho, \vartheta)
\ &= \ 
\frac{1}{H_{\ell}}
\left(
\begin{array}{cccc}
K_{\ell} & -V_{\ell} & 0 & 0 
\\
-V_{\ell} & 1 & 0 & 0 
\\
0 & 0 & 1 & V_{\ell} 
\\ 
0 & 0 & V_{\ell} & K_{\ell} 
\end{array}
\right)
\, , \ls 
\Omega^{\text{KK}}
\ = \ 
\left(
\begin{array}{cccc}
1 & 0 & 0 & 0 
\\
-2 \pi \ell & 1 & 0 & 0 
\\
0 & 0 & 1 & 2 \pi \ell
\\ 
0 & 0 & 0 & 1
\end{array}
\right)
\, . 
\label{SO22-MM-dKK5}
\end{align}
The monodromy matrix $\Omega^{\text{KK}}$ does not contain the $\Theta$ part and the $\beta$ part.
This is also consistent with the configuration (\ref{dKK5-system}),
where the $2 \pi$ shift of the coordinate $\vartheta$ can be eliminated by the coordinate transformations.
However, the matrices of the exotic $5^2_2$-brane (\ref{522-system}),
\begin{align}
\mathscr{M}^{\text{E}} (\varrho, \vartheta)
\ &= \ 
\frac{1}{H_{\ell}}
\left(
\begin{array}{cccc}
1 & 0 & 0 & - V_{\ell} 
\\
0 & 1 & V_{\ell} & 0 
\\
0 & V_{\ell} & K_{\ell} & 0 
\\
- V_{\ell} & 0 & 0 & K_{\ell}
\end{array}
\right)
\, , \ls 
\Omega^{\text{E}}
\ = \ 
\left(
\begin{array}{cccc}
1 & 0 & 0 & - 2 \pi \ell
\\
0 & 1 & 2 \pi \ell & 0 
\\
0 & 0 & 1 & 0 
\\
0 & 0 & 0 & 1
\end{array}
\right)
\, , 
\label{SO22-MM-522}
\end{align}
indicates the exotic feature because
the monodromy matrix $\Omega^{\text{E}}$ contains the $\beta$ part.
Indeed, in the configuration (\ref{522-system}), 
the $2 \pi$ shift of the coordinate $\vartheta$ cannot be removed by the coordinate transformations and the B-field gauge transformation.
This shift is generated by the T-duality symmetry along the two-torus $T^{89}$.
Hence we can interpret that the background geometry of the exotic $5^2_2$-brane is a typical example of T-folds \cite{Hull:2004in}.

Monodromy is quite useful to investigate (non)geometric aspects.
Furthermore, if we apply the equivalence $O(2,2;{\mathbb Z}) = SL(2,{\mathbb Z}) \times SL(2,{\mathbb Z})$ to 
the analysis of monodromy,
we can explore the geometries of defect five-branes in a deeper level.
In the next two sections we will carefully analyze the $SL(2,{\mathbb Z}) \times SL(2,{\mathbb Z})$ monodromy and construct new configurations of defect five-branes.

\section{$SL(2,{\mathbb Z}) \times SL(2,{\mathbb Z})$ monodromy of defect five-branes}
\label{sect:SL2SL2-monodromy}

In the previous section we studied the defect five-branes and their $O(2,2;{\mathbb Z})$ monodromy structures.
In this section we apply the equivalence
$O(2,2;{\mathbb Z}) = SL(2,{\mathbb Z}) \times SL(2,{\mathbb Z})$ to 
monodromy of the defect five-branes \cite{deBoer:2012ma}.

\subsection{Two complex structures}

The $O(2,2;{\mathbb Z})$ monodromy is generated by the two-torus $T^{89}$.
Let us further study the monodromy by the equivalent group $SL(2,{\mathbb Z}) \times SL(2,{\mathbb Z})$.
Each $SL(2,{\mathbb Z})$ should also be governed by the structure of $T^{89}$.
Associated with these two $SL(2,{\mathbb Z})$, we introduce two complex structures $\tau$ and $\rho$.
$\tau$ is the complex structure of the two-torus $T^{89}$,
while $\rho$ is defined in terms of the B-field and the metric on $T^{89}$
in such a way as \cite{deBoer:2012ma},
\bsubeq \label{field-config}
\begin{align}
\rho \ &\equiv \ 
B_{89} + \I \sqrt{\det G_{mn}}
\, .
\end{align}
In terms of the two complex structures,
we can represent the metric $G_{mn}$ and the B-field $B_{mn}$ on the two-torus $T^{89}$, and the dilaton $\phi$, 
\begin{gather}
G_{mn} \ = \ 
\frac{\rho_2}{\tau_2} \left(
\begin{array}{cc}
1 & \tau_1
\\
\tau_1 & |\tau|^2
\end{array}
\right)
\, , \ls
B_{89} \ = \ \rho_1
\, , \ls
\e^{2 \phi} \ = \ \rho_2
\, , 
\end{gather}
\esubeq
where $\tau = \tau_1 + \I \tau_2$ and $\rho = \rho_1 + \I \rho_2$.
Then, instead of the analysis of the matrix $\mathscr{M}(\varrho, \vartheta)$,
we will investigate the monodromy structures of the two complex structures $\tau$ and $\rho$ of the defect five-branes.
The configuration of the dilaton is also fixed in order to satisfy the equations of motion of supergravity theories \cite{deBoer:2012ma}.

\subsection{Monodromy matrices}


Let us first analyze the $SL(2,{\mathbb Z})_{\tau} \times SL(2,{\mathbb Z})_{\rho}$ monodromy of the defect NS5-brane. 
Plugging the configuration (\ref{dNS5-system}) into the formulation (\ref{field-config}), 
we can read off the explicit forms of the two complex structures,
\begin{align}
\tau \ &= \ \I
\, , \ls
\rho \ = \
V_{\ell} + \I H_{\ell}
\ = \ 
\I h + \I \ell \, \log (\mu /z) 
\, , \label{original-taurho-dNS5}
\end{align}
where we defined the complex coordinate $z \equiv \varrho \, \e^{\I \vartheta}$ in the 67-plane.
When we go around the defect NS5-brane $z \to z \, \e^{2 \pi \I}$,
the complex structure $\rho$ has the monodromy as $\rho \ \to \ \rho + 2 \pi \ell$,
whilst $\tau$ is invariant.
The $SL(2,{\mathbb Z})$ descriptions of the monodromy are 
\bsubeq \label{dNS5-SL2SL2-monodromy}
\begin{alignat}{2}
\tau \ &\to \ 
\tau' \ = \ \tau 
\, , &\ls
\Omega_{\tau}^{\text{NS}}
\ &\equiv \ 
\left(
\begin{array}{cc}
1 & 0 \\
0 & 1
\end{array}
\right)
\, , \\
\rho \ &\to \ 
\rho' \ = \ \rho + 2 \pi \ell 
\, , &\ls
\Omega_{\rho}^{\text{NS}}
\ &\equiv \ 
\left(
\begin{array}{cc}
1 & 2 \pi \ell \\
0 & 1
\end{array}
\right)
\, .
\end{alignat}
\esubeq
It turns out that the two-torus $T^{89}$ is not deformed under the monodromy,
while the field configuration of $B_{89}$ is changed.
However, this change can be removed by the B-field gauge transformation.
Then the configuration (\ref{dNS5-system}) is invariant under the monodromy transformation.
This is consistent with the previous analysis in terms of $\Omega^{\text{NS}}$ (\ref{SO22-MM-dNS5}).

Next, we discuss the $SL(2,{\mathbb Z})_{\tau} \times SL(2,{\mathbb Z})_{\rho}$ monodromy of the defect KK5-brane.
Substituting the configuration (\ref{dKK5-system}) into (\ref{field-config}),
the two complex structures are given as
\begin{align}
\tau \ &= \ 
\frac{- V_{\ell} + \I H_{\ell}}{K_{\ell}}
\ = \ 
\frac{\I}{h + \ell \log (\mu/z)}
\, , \ls
\rho \ = \ \I
\, . \label{original-taurho-dKK5}
\end{align}
Now $\rho$ becomes trivial.
Here it is convenient to introduce $\lambda = -1/\tau = V_{\ell} + \I H_{\ell}$.
Under the shift $z \to z \, \e^{2 \pi \I}$,
the complex structure $\lambda$ is transformed as $\lambda \to \lambda' = \lambda + 2 \pi \ell$, while $\rho$ is invariant. 
Their $SL(2,{\mathbb Z})$ representations are given as follows,
\bsubeq \label{dKK5-SL2SL2-monodromy}
\begin{alignat}{2}
\tau \ &\to \ 
\tau' \ = \ \frac{\tau}{- 2 \pi \ell \tau + 1}
\, , &\ls
\Omega_{\tau}^{\text{KK}}
\ &\equiv \ 
\left(
\begin{array}{cc}
1 & 0 \\
- 2 \pi \ell & 1
\end{array}
\right)
\, , \\
\rho \ &\to \ 
\rho' \ = \ \rho 
\, , &\ls
\Omega_{\rho}^{\text{KK}}
\ &\equiv \ 
\left(
\begin{array}{cc}
1 & 0 \\
0 & 1
\end{array}
\right)
\, .
\end{alignat}
\esubeq
This implies that the complex structure of the two-torus $T^{89}$ is changed under the monodromy, 
while the B-field and the determinant of the metric is invariant.
However, we can remove the change of the complex structure by the coordinate transformations.
This is also consistent with the previous discussion in terms of $\Omega^{\text{KK}}$ (\ref{SO22-MM-dKK5}).


Finally, we study the $SL(2,{\mathbb Z})_{\tau} \times SL(2,{\mathbb Z})_{\rho}$ monodromy of the exotic $5^2_2$-brane.
Applying the configuration (\ref{522-system}) to the complex structures (\ref{field-config}),
we can read off the following forms,
\begin{align}
\tau \ &= \ \I
\, , \ls
\rho \ = \ 
\frac{- V_{\ell} + \I H_{\ell}}{K_{\ell}}
\ = \ 
\frac{\I}{h + \ell \log (\mu/z)}
\, . \label{original-taurho-522}
\end{align}
Again the complex structure of the two-torus $T^{89}$ becomes trivial.
For convenience, 
we introduce $\omega \equiv -1/\rho = V_{\ell} + \I H_{\ell}$.
Under the shift $z \to z \, \e^{2 \pi \I}$, 
we see that $\omega$ has the monodromy 
$\omega \to \omega' = \omega + 2 \pi \ell$,
while $\tau$ is invariant.
The $SL(2,{\mathbb Z})$ matrix forms of the monodromy are 
\bsubeq \label{522-SL2SL2-monodromy}
\begin{alignat}{2}
\tau \ &\to \ 
\tau' \ = \ \tau 
\, , &\ls
\Omega_{\tau}^{\text{E}}
\ &\equiv \ 
\left(
\begin{array}{cc}
1 & 0 \\
0 & 1
\end{array}
\right)
\, , \\
\rho \ &\to \ 
\rho' \ = \ \frac{\rho}{- 2 \pi \ell \rho + 1}
\, , &\ls
\Omega_{\rho}^{\text{E}}
\ &\equiv \ 
\left(
\begin{array}{cc}
1 & 0 \\
- 2 \pi \ell & 1
\end{array}
\right)
\, .
\end{alignat}
\esubeq
This behavior implies that the monodromy transformation does not change the complex structure of the two-torus, while the field configuration is changed.
Furthermore, caused by the form $\rho' = B'_{89} + \I \sqrt{\det G'_{mn}}$, 
this change cannot be eliminated completely in terms of the coordinate transformations and the B-field gauge transformation.
As discussed in the analysis of $\Omega^{\text{E}}$, 
this is nothing but the aspect of T-fold.

In the next section we will investigate conjugates of 
$SL(2,{\mathbb Z})_{\tau} \times SL(2,{\mathbb Z})_{\rho}$ monodromy.
We will find various new configurations of defect five-branes as their composite states.

\section{Conjugate configurations}
\label{sect:ConjugateConfig}

In type IIB theory,
there exists a D7-brane which also has $SL(2,{\mathbb Z})$ monodromy generated by the combination of the dilaton and the axion \cite{Greene:1989ya}.
Applying a generic $SL(2,{\mathbb Z})$ transformation to the monodromy,
we can find a conjugate system of the D7-brane \cite{DeWolfe:1998eu, DeWolfe:1998pr}.
In the same analogy, a conjugate of an exotic $5^2_2$-brane has been discussed in \cite{Kikuchi:2012za, deBoer:2012ma}.
In this section, we develop the analyses to conjugate configurations of the defect five-branes.
To do this, we prepare a set of generic $SL(2,{\mathbb Z})_{\tau} \times SL(2,{\mathbb Z})_{\rho}$ matrices,
\bsubeq \label{conjugate-U}
\begin{gather}
U_{\tau} \ \equiv \ 
\left(
\begin{array}{cc}
s' & r' \\
q' & p'
\end{array}
\right)
\, , \ls
U_{\rho} \ \equiv \ 
\left(
\begin{array}{cc}
s & r \\
q & p
\end{array}
\right)
\, , \ls
\begin{array}{rcl}
s'p' - r'q' \!\!&=&\!\! 1
\, , \\
sp - rq \!\!&=&\!\! 1
\, .
\end{array}
\end{gather}
By using $U_{\tau,\rho}$, we construct a set of conjugate monodromy matrices $\wt{\Omega}_{\tau,\rho}$,
\begin{align}
\Omega_{\tau,\rho} \ \to \ 
\wt{\Omega}_{\tau,\rho} \ = \ 
U_{\tau,\rho}^{-1} \Omega_{\tau,\rho}^{\vphantom{-1}} U_{\tau,\rho}^{\vphantom{-1}}
\, .
\end{align}
\esubeq
Simultaneously, we transform the two complex structures $\tau$ and $\rho$ in terms of $U_{\tau,\rho}$ to new complex structures $\wt{\tau}$ and $\wt{\rho}$.
Plugging them into (\ref{field-config}), we can read of new field configurations $\wt{G}_{mn}$, $\wt{B}_{89}$ and $\wt{\phi}$.

\subsection{Conjugate configuration of defect NS5-brane}
\label{cc-dNS5}

First, we investigate a conjugate configuration
of the defect NS5-brane (\ref{dNS5-system}).
Transforming the original monodromy matrices $\Omega_{\tau}^{\text{NS}}$ and $\Omega_{\rho}^{\text{NS}}$ (\ref{dNS5-SL2SL2-monodromy}) in terms of the rule (\ref{conjugate-U}),
the conjugate monodromy matrices can be obtained as
\begin{align}
\wt{\Omega}_{\tau}^{\text{NS}}
\ &= \ 
\left(
\begin{array}{cc}
1 & 0 \\
0 & 1
\end{array}
\right)
\, , \ls
\wt{\Omega}_{\rho}^{\text{NS}}
\ = \ 
\left(
\begin{array}{cc}
1 + 2 \pi \ell p q & 2 \pi \ell p^2 
\\
- 2 \pi \ell q^2 & 1 - 2 \pi \ell pq
\end{array}
\right)
\, . \label{conjugate-dNS5}
\end{align}
Here $\wt{\Omega}_{\tau}^{\text{NS}}$ is again trivial because the original complex structure $\tau$ is trivial $\tau = \I$.
Compared $\wt{\Omega}_{\rho}^{\text{NS}}$ with $\Omega_{\rho}^{\text{NS}}$ (\ref{dNS5-SL2SL2-monodromy}) and $\Omega_{\rho}^{\text{E}}$ (\ref{522-SL2SL2-monodromy}),
it turns out that 
the conjugate system is a composite of $p$ defect NS5-branes and $q$ exotic $5^2_2$-branes \cite{Kikuchi:2012za, deBoer:2012ma}.
Associated with the transformation rule (\ref{conjugate-U}),
we also arrange the complex structure $\rho$ by means of $U_{\rho}^{-1}$,
\begin{align}
U_{\rho}^{-1} \ = \ 
\left(
\begin{array}{cc}
p & -r \\
- q & s
\end{array}
\right)
\, , \ls
\rho \ \to \ 
\wt{\rho} \ &\equiv \
\frac{p \rho - r}{-q \rho + s}
\, . \label{conjugate-rho-dNS5}
\end{align}

Since the monodromy of the original $\rho$ is 
$\rho \to \rho' = \rho + 2 \pi \ell$ under the shift $z \to z \, \e^{2 \pi \I}$,
then the new complex structure $\wt{\rho}$ is transformed as
\begin{align}
\wt{\rho} \ \to \ \wt{\rho}'
\ &= \
\frac{p \rho' - r}{-q \rho' + s}
\ = \
\frac{(1 + 2 \pi \ell pq) \wt{\rho} + 2 \pi \ell p^2}{- 2 \pi \ell q^2 \wt{\rho} + (1 - 2 \pi \ell pq)}
\, . 
\end{align}
This indicates that the new complex structure $\wt{\rho}$ reproduces the $SL(2,{\mathbb Z})_{\rho}$ conjugate monodromy $\wt{\Omega}_{\rho}^{\text{NS}}$ (\ref{conjugate-dNS5}).
Then it turns out that $\wt{\rho}$ denotes a conjugate configuration of the defect NS5-brane (\ref{dNS5-system}).

We would like to construct a local expression of the conjugate configuration.
Substituting the conjugate complex structures $\tau$ and $\wt{\rho}$ (\ref{conjugate-rho-dNS5}) into (\ref{field-config}), 
we can read off the field configuration,
\bsubeq \label{conjugate-BGphi-dNS5}
\begin{align}
\wt{G}_{mn} \ &= \ 
\frac{H_{\ell}}{s^2 - 2 qs V_{\ell} + q^2 K_{\ell}} \, \delta_{mn}
\, , \ls
m,n = 8,9
\, , \\
\wt{B}_{89} \ &= \ 
\frac{ -rs + (ps + qr) V_{\ell} - pq K_{\ell}}{s^2 - 2 qs V_{\ell} + q^2 K_{\ell}}
\, , \\
\e^{2 \wt{\phi}} 
\ &= \ 
\frac{H_{\ell}}{s^2 - 2 qs V_{\ell} + q^2 K_{\ell}}
\, .
\end{align}
\esubeq
This is a generic form of a composite system of $p$ defect NS5-branes and $q$ exotic $5^2_2$-branes under the constraint $sp-qr = 1$.
The system of a single defect NS5-brane can be realized by setting $(p,q,r,s) = (1,0,0,1)$, 
while the system of a single exotic $5^2_2$-brane can be expressed by $(p,q,r,s) = (0,1,-1,0)$.

In a generic case of non-vanishing $p$ and $q$, the expression (\ref{conjugate-BGphi-dNS5}) is rather lengthy.
In order to reduce the expression of the generic $(p,q)$ configuration,
we rewrite the conjugate complex structure $\wt{\rho}$,
\begin{align}
\wt{\rho}
\ &= \ 
\frac{p(V_{\ell} + \I H_{\ell}) - r}{-q (V_{\ell} + \I H_{\ell}) + s}
\ = \ 
- \frac{p}{q} - \frac{1}{q [(-s + q V_{\ell}) + \I q H_{\ell}]}
\, . \label{rho2-dNS5}
\end{align}
Here we removed $r$ by using $sp-qr= 1$. 
Now $s$ is no longer constrained by $(p,q)$,
then we set $s$ to zero without loss of generality,
\begin{align}
\wh{\rho} \ &\equiv \ 
- \frac{p}{q} - \frac{1}{q^2 (V_{\ell} + \I H_{\ell})}
\, . \label{rho3-dNS5}
\end{align}
This reduction can be interpreted as the coordinate transformation of the angular coordinate $\vartheta$.
For convenience, we further introduce a new expression $\wh{\omega} \equiv -1/(\wh{\rho} + \frac{p}{q})$.
This is transformed as $\wh{\omega} \to \wh{\omega}' = \wh{\omega} + 2 \pi \ell q^2$ under the shift $z \to z\,\e^{2 \pi \I}$.
Then we can read off 
the monodromy of $\wh{\rho} = -1/\wh{\omega}$ in such a way as 
\begin{align}
\wh{\rho} \ \to \ 
\wh{\rho}' \ &= \ 
\frac{(1 + 2 \pi \ell pq) \wh{\rho} + 2 \pi \ell p^2}{- 2 \pi \ell q^2 \wh{\rho} + (1 - 2 \pi \ell pq)}
\, .
\end{align}
It turns out that the reduced complex structure $\wh{\rho}$ again reproduces
the conjugate monodromy matrix $\wt{\Omega}_{\rho}^{\text{NS}}$ (\ref{conjugate-rho-dNS5}).
Throughout the above reduction, the other complex structure $\tau$ is unchanged. 
Finally, plugging $\tau = \I$ and $\wh{\rho}$ (\ref{rho3-dNS5}) into the definition (\ref{field-config}), 
we explicitly obtain the local description of a defect $(p,q)$ five-brane,
i.e, the configuration of $p$ defect NS5-branes and $q$ exotic $5^2_2$-branes, 
\bsubeq \label{ConjugateConfig-dNS5}
\begin{gather}
\d s^2 \ = \ 
\d s_{012345}^2 
+ H_{\ell} \big[ (\d \varrho)^2 + \varrho^2 (\d \vartheta)^2 \big]
+ \frac{H_{\ell}}{q^2 K_{\ell}} \big[ (\d x^8)^2 + (\d x^9)^2 \big]
\, , \\
\wh{B}_{89} \ = \ 
- \frac{p}{q} - \frac{V_{\ell}}{q^2 K_{\ell}}
\, , \ls
\e^{2 \wh{\phi}} \ = \ 
\frac{H_{\ell}}{q^2 K_{\ell}} 
\, .
\end{gather}
\esubeq
Of course, this configuration satisfies the equations of motion of supergravity theories.
In the previous work \cite{Kimura:2013khz}, we could not find the local description (\ref{ConjugateConfig-dNS5}).
At that time we did not have any ideas how to use the monodromy transformation in a suitable way to describe a composite of defect five-branes. 
The description (\ref{ConjugateConfig-dNS5}) would enable us to construct a correct GLSM for a defect $(p,q)$ five-brane. 
Thus the configuration (\ref{ConjugateConfig-dNS5}) is indeed the one which we wanted to describe in \cite{Kimura:2013khz}.

We argue the mass of the defect $(p,q)$ five-brane (\ref{ConjugateConfig-dNS5}). 
In this case the metric and the dilaton are quite similar to those of the single exotic $5^2_2$-brane (\ref{522-system}).
The integer $p$, the number of defect NS5-branes, only appears in the constant term of the B-field (\ref{ConjugateConfig-dNS5}).
Since the mass of defect five-branes can be evaluated in terms of the metric and the dilaton under a certain assumption (see the section 4 of \cite{deBoer:2012ma}),
we guess that the mass of $p$ defect NS5-branes in the system (\ref{ConjugateConfig-dNS5}) is negligible.
This is indeed true, as far as we concern the supergravity regime where the dilaton is very small
\begin{align}
g_{\text{st}}^2 \ = \ \e^{2 \wh{\phi}} \ = \ \frac{H_{\ell}}{q^2 K_{\ell}}
\ \ll \ 1
\, . 
\end{align}
Now the system (\ref{ConjugateConfig-dNS5}) has the physical coordinates $x^{8,9}$ whose radii are $R_{8,9}$.
They are related to the function $H_{\ell}/(q^2 K_{\ell})$,
\begin{align}
\Big( \frac{R_8}{\ell_{\text{st}}} \Big)^2 \ = \ 
\Big( \frac{R_9}{\ell_{\text{st}}} \Big)^2 \ = \ 
\frac{H_{\ell}}{q^2 K_{\ell}}
\, , 
\end{align}
in the appropriately large $\varrho$ region.
This implies that the parameters $R_{8,9}/\ell_{\text{st}}$ are very small in the supergravity regime.
In terms of the radii of the physical coordinates of (\ref{ConjugateConfig-dNS5}), we can evaluate the mass of a defect NS5-branes (\ref{massdNS5}) and an exotic $5^2_2$-brane (\ref{mass522}) in such a way as
\begin{align}
M_{\text{NS}} \ &= \ 
\frac{1}{g_{\text{st}}^2 \ell_{\text{st}}^6}
\, , \ls
M_{\text{E}} \ = \ 
\frac{1}{g_{\text{st}}^2 \ell_{\text{st}}^2 (R_8 R_9)^2}
\, , \ls
\frac{M_{\text{NS}}}{M_{\text{E}}}
\ = \ 
\frac{(R_8 R_9)^2}{\ell_{\text{st}}^4}
\ = \ \Big( \frac{H_{\ell}}{q^2 K_{\ell}} \Big)^2
\ \ll \ 1
\, . 
\end{align}
Hence it turns out that the mass of $p$ defect NS5-branes in the composite system (\ref{ConjugateConfig-dNS5}) is negligible.

There is a comment.
The conjugate complex structure $\wh{\rho}$ (\ref{rho3-dNS5}) contains a constant term $- p/q$.
Compared this with (\ref{field-config}),
we think that this constant might be eliminated by the B-field gauge symmetry.
However, 
if the term $- p/q$ in (\ref{rho3-dNS5}) is gauged away, 
the monodromy matrix $\wt{\Omega}_{\rho}^{\text{NS}}$ is reduced to $(\Omega_{\rho}^{\text{E}})^q$, 
which no longer represents a composite system of defect NS5-branes and exotic $5^2_2$-branes.

\subsection{Conjugate configuration of defect KK5-brane}

Next, we consider a conjugate of the defect KK5-brane (\ref{dKK5-system}).
Applying the transformation rules (\ref{conjugate-U}) to the monodromy matrices
$\Omega_{\tau}^{\text{KK}}$ and $\Omega_{\rho}^{\text{KK}}$ (\ref{dKK5-SL2SL2-monodromy}), 
the conjugate monodromy matrices are given as
\begin{align}
\wt{\Omega}_{\tau}^{\text{KK}}
\ &= \ 
\left(
\begin{array}{cc}
1 + 2 \pi \ell r's' & 2 \pi \ell r'{}^2 
\\
- 2 \pi \ell s'{}^2 & 1 - 2 \pi \ell r's'
\end{array}
\right)
\, , \ls
\wt{\Omega}_{\rho}^{\text{KK}}
\ = \ 
\left(
\begin{array}{cc}
1 & 0 \\
0 & 1
\end{array}
\right)
\, . \label{conjugate-dKK5}
\end{align}
We note that the conjugate monodromy matrix
$\wt{\Omega}_{\rho}^{\text{KK}}$ is identical with $\Omega_{\rho}^{\text{KK}}$ because the complex structure $\rho$ is trivial (\ref{original-taurho-dKK5}).
Compared $\wt{\Omega}_{\tau}^{\text{KK}}$ with the monodromy matrices 
$\Omega_{\tau}^{\text{KK}}$ (\ref{dKK5-SL2SL2-monodromy})
and $\Omega_{\tau}^{\text{AK}}$ (\ref{adKK5-SL2SL2-monodromy}),
the conjugate monodromy matrix
$\wt{\Omega}_{\tau}^{\text{KK}}$ denotes that the system is a composite of $-s'$ defect KK5-branes (\ref{dKK5-system}) and $r'$ defect KK5-branes of another type (\ref{adKK5-system}). 
Let us focus on the complex structure $\tau$.
This is also changed in terms of $U_{\tau}^{-1}$ in such a way as
\begin{align}
U_{\tau}^{-1} \ = \ 
\left(
\begin{array}{cc}
p' & -r' \\
-q' & s'
\end{array}
\right)
\, , \ls
\tau \ \to \ 
\wt{\tau} \ &\equiv \
\frac{p' \tau - r'}{-q' \tau + s'}
\, . \label{conjugate-tau-dKK5}
\end{align}
Since the original $\tau$ is transformed as in (\ref{dKK5-SL2SL2-monodromy})
under the shift $z \to z \, \e^{2 \pi \I}$,
the new complex structure $\wt{\tau}$ is transformed,
\begin{align}
\wt{\tau} \ \to \ 
\wt{\tau}' \ &= \ 
\frac{(1 + 2 \pi \ell r' s') \wt{\tau} + 2 \pi \ell r'{}^2}{- 2 \pi \ell s'{}^2 \wt{\tau} + (1 - 2 \pi \ell r' s')}
\, .
\end{align}
This provides the same conjugate monodromy matrix $\wt{\Omega}_{\tau}^{\text{KK}}$ (\ref{conjugate-dKK5}).
Plugging $\wt{\tau}$ (\ref{conjugate-tau-dKK5}) and $\rho = \I$ into the definition (\ref{field-config}), we find
\bsubeq \label{conjugate-BGphi-dKK5}
\begin{align}
\wt{G}_{88} 
\ &= \ 
\frac{q'{}^2 + 2 q's' V_{\ell} + s'{}^2 K_{\ell}}{H_{\ell}}
\, , \\
\wt{G}_{89} \ = \ \wt{G}_{98}
\ &= \ 
- \frac{p'q' + (p's' + r'q') V_{\ell} + r' s' K_{\ell}}{H_{\ell}}
\, , \\
\wt{G}_{99} 
\ &= \ 
\frac{p'{}^2 + 2 p'r' V_{\ell} + r'{}^2 K_{\ell}}{H_{\ell}}
\, , \\
\wt{B}_{MN} \ &= \ 0
\, , \ls
\e^{2 \wt{\phi}} \ = \ 1
\, .
\end{align}
\esubeq
This is a generic form of a composite system of $-s'$ defect KK5-branes and $r'$ defect KK5-branes of another type.
The system of the single defect KK5-brane (\ref{dKK5-system}) is obtained by setting $(p',q',r',s') = (-1,0,0,-1)$,
while the configuration of the single defect KK5-brane of another type (\ref{adKK5-system}) is realized if $(p',q',r',s') = (0,-1,1,0)$.

The expression (\ref{conjugate-BGphi-dKK5}) is lengthy to describe the generic configuration of non-vanishing parameters $(-s',r')$.
Let us find the simple form of the generic $(-s',r')$, 
We introduce $\wt{\lambda} = -1/\wt{\tau}$.
Applying the constraint $s'p'-q'r' = 1$ to this, we obtain 
\bsubeq \label{tau2-dKK5}
\begin{align}
\wt{\lambda} 
\ &= \ 
\frac{s' (V_{\ell} + \I H_{\ell}) + q'}{r' (V_{\ell} + \I H_{\ell}) + p'}
\ = \ 
\frac{s'}{r'} - \frac{1}{r' [(p' + r' V_{\ell}) + \I r' H_{\ell}]}
\, .
\end{align}
\esubeq
Now the parameter $p'$ is no longer constrained by the other parameters $(-s',r')$.
Then we can set $p' = 0$ by the coordinate transformation of $\vartheta$,
\begin{align}
\wh{\lambda} \ = \ 
- \frac{1}{\wh{\tau}}
\ &\equiv \ 
\frac{s'}{r'} - \frac{1}{r'{}^2 (V_{\ell} + \I H_{\ell})}
\, . \label{tau3-dKK5}
\end{align}
We check the monodromy of the new complex structure $\wh{\tau}$.
It is convenient to introduce $\wh{\zeta} \equiv -1/(\wh{\lambda} - \frac{s'}{r'})$.
Since this is transformed as 
$\wh{\zeta} \to \wh{\zeta}' = \wh{\zeta} + 2 \pi \ell r'{}^2$ 
under the shift $z \to z\,\e^{2 \pi \I}$,
we can immediately read off the transformation of $\wh{\tau} = - 1/\wh{\lambda}$ in such a way as
\begin{align}
\wh{\tau} \ \to \ 
\wh{\tau}' \ &= \
\frac{(1 + 2 \pi \ell r's') \wh{\tau} + 2 \pi \ell r'{}^2}{- 2 \pi \ell s'{}^2 \wh{\tau} + (1 - 2 \pi \ell r's')}
\, .
\end{align}
This guarantees that the new complex structure $\wh{\tau}$ also generates the conjugate monodromy (\ref{conjugate-dKK5}).
Plugging $\wh{\tau}$ (\ref{tau3-dKK5}) and $\rho = \I$ into the definition (\ref{field-config}), 
we find the local expression of the metric, the B-field and the dilaton 
for the composite of $-s'$ defect KK5-branes (\ref{dKK5-system}) and $r'$ defect KK5-branes of another type (\ref{adKK5-system}),
\bsubeq \label{ConjugateConfig-dKK5}
\begin{gather}
\d s^2 \ = \ 
\d s_{012345}^2 
+ H_{\ell} \big[ (\d \varrho)^2 + \varrho^2 (\d \vartheta)^2 \big]
+ \wh{\lambda}_2 \, (\d x^8)^2
+ \frac{1}{\wh{\lambda}_2} \big[ \d y^9 - \wh{\lambda}_1 \, \d x^8 \big]^2
\, , \\
\wh{B}_{MN} \ = \ 0
\, , \ls
\e^{2 \wh{\phi}} \ = \ 1
\, , \\
\wh{\lambda}_1 \ = \
- \frac{\wh{\tau}_1}{|\wh{\tau}|^2}
\ = \ 
\frac{s'}{r'} - \frac{V_{\ell}}{r'{}^2 K_{\ell}}
\, , \ls
\wh{\lambda}_2 \ = \ 
\frac{\wh{\tau}_2}{|\wh{\tau}|^2}
\ = \ 
\frac{H_{\ell}}{r'{}^2 K_{\ell}}
\, .
\end{gather}
\esubeq
This configuration also satisfies the equations of motion of supergravity theories.
We note that the transverse space of 6789-directions in (\ref{ConjugateConfig-dKK5}) is Ricci-flat.
Since this configuration preserves a half of supersymmetry,
the transverse space is also hyper-K\"{a}hler.
Indeed this belongs to a class of ALG spaces \cite{Cherkis:2000cj, Cherkis:2001gm}\footnote{The transverse space of the defect KK5-brane (\ref{dKK5-system}) is also an ALG space.}.

The conjugate system can also be obtained from the conjugate configuration of the defect NS5-brane (\ref{ConjugateConfig-dNS5}) via the T-duality transformation along the 9-th direction with relabeling $(p,q) = (-s',r')$.
The exchange of the conjugate complex structures $(\tau, \wh{\rho})$ in (\ref{ConjugateConfig-dNS5}) and $(\rho, \wh{\lambda})$ in (\ref{ConjugateConfig-dKK5}) also occurs.

We notice that 
the constant term $s'/r'$ in the conjugate complex structure $\wh{\tau}$ (\ref{tau3-dKK5}) should not be eliminated in terms of the coordinate transformations.
If we remove $s'/r'$, the complex structure produces the monodromy matrix
$(\Omega_{\tau}^{\text{AK}})^{r'}$ (\ref{adKK5-SL2SL2-monodromy}) rather than the conjugate monodromy matrix $\wt{\Omega}_{\tau}^{\text{KK}}$ (\ref{conjugate-dKK5}).

\subsection{Conjugate configuration of exotic $5^2_2$-brane}

Finally, we investigate a conjugate system of the exotic $5^2_2$-brane (\ref{522-system}).
The $SL(2,{\mathbb Z})_{\tau} \times SL(2,{\mathbb Z})_{\rho}$ monodromy matrices (\ref{522-SL2SL2-monodromy}) are transformed by using (\ref{conjugate-U}),
\begin{align}
\wt{\Omega}_{\tau}^{\text{E}}
\ &= \ 
\left(
\begin{array}{cc}
1 & 0 \\
0 & 1
\end{array}
\right)
\, , \ls
\wt{\Omega}_{\rho}^{\text{E}}
\ = \ 
\left(
\begin{array}{cc}
1 + 2 \pi \ell rs & 2 \pi \ell r^2 
\\
- 2 \pi \ell s^2 & 1 - 2 \pi \ell rs
\end{array}
\right)
\, . \label{conjugate-522}
\end{align}
Since the complex structure $\tau$ (\ref{original-taurho-522}) is trivial,
the conjugate monodromy matrix
$\wt{\Omega}_{\tau}^{\text{E}}$ coincides with the original one $\Omega_{\tau}^{\text{E}}$.
On the other hand, the conjugate monodromy matrix
$\wt{\Omega}_{\rho}^{\text{E}}$ implies that the conjugate system consists of $r$ defect NS5-branes and $-s$ exotic $5^2_2$-branes.
In order to obtain an explicit field configuration of the conjugate system,
we also transform the complex structure $\rho$,
\begin{align}
U_{\rho}^{-1} \ = \ 
\left(
\begin{array}{cc}
p & -r \\
-q & s
\end{array}
\right)
\, , \ls
\rho \ \to \ 
\wt{\rho} \ &\equiv \
\frac{p \rho - r}{-q \rho + s}
\, . \label{conjugate-rho-522}
\end{align}
Recall that the original $\rho$ is transformed as in (\ref{522-SL2SL2-monodromy})
under $z \to z \, \e^{2 \pi \I}$.
Then the new complex structure $\wt{\rho}$ is also transformed as
\begin{align}
\wt{\rho} \ \to \ 
\wt{\rho}' \ &= \ 
\frac{(1 + 2 \pi \ell rs) \wt{\rho} + 2 \pi \ell r^2}{- 2 \pi \ell s^2 \wt{\rho} + (1 - 2 \pi \ell rs)}
\, .
\end{align}
This reproduces the conjugate monodromy matrix
$\wt{\Omega}_{\rho}^{\text{E}}$ (\ref{conjugate-522}).
The complex structures $\tau$ and $\wt{\rho}$ gives rise to a conjugate configuration of the metric, the B-field and the dilaton,
\bsubeq \label{conjugate-BGphi-522}
\begin{align}
\wt{G}_{mn} \ &= \ 
\frac{H_{\ell}}{q^2 + 2 qs V_{\ell} + s^2 K_{\ell}}
\, \delta_{mn}
\, , \ls
m,n = 8,9
\, , \\
\wt{B}_{89} \ &= \ 
- \frac{pq + (ps + rq) V_{\ell} + rs K_{\ell}}{q^2 + 2 qs V_{\ell} + s^2 K_{\ell}}
\, , \\
\e^{2 \wt{\phi}} \ &= \ 
\frac{H_{\ell}}{q^2 + 2 qs V_{\ell} + s^2 K_{\ell}}
\, .
\end{align}
\esubeq
This contains the case of the single exotic $5^2_2$-brane (\ref{522-system}) by setting $(p,q,r,s) = (-1,0,0,-1)$ and 
the case of the single defect NS5-brane (\ref{dNS5-system})
by $(p,q,r,s) = (0,-1,1,0)$.
However, in the case of a generic $(r,-s) \neq (0,0)$,
the expression (\ref{conjugate-BGphi-522}) is cumbersome.
Fortunately, as in the previous configurations, 
we can reduce (\ref{conjugate-BGphi-522}). 
For convenience, let us introduce $\wt{\omega} = -1/\wt{\rho}$,
\bsubeq \label{rho2-522}
\begin{align}
\wt{\omega} 
\ &= \ 
\frac{s (V_{\ell} + \I H_{\ell}) + q}{r (V_{\ell} + \I H_{\ell}) + p}
\ = \
\frac{s}{r} - \frac{1}{r [(p + r V_{\ell}) + \I r H_{\ell}]}
\, ,
\end{align}
\esubeq
where we used $sp - qr = 1$ to remove $q$.
Since the parameter $p$ is now arbitrary without any constraints,
we can set the following form by the coordinate transformation of $\vartheta$,
\begin{align}
\wh{\omega} \ = \ - \frac{1}{\wh{\rho}}
\ &\equiv \ 
\frac{s}{r} - \frac{1}{r^2 (V_{\ell} + \I H_{\ell})}
\, . \label{rho3-522}
\end{align}
We check the monodromy of the new complex structure $\wh{\rho}$.
For convenience, we define $\wh{\zeta} \equiv -1/(\wh{\omega} - \frac{s}{r})$.
This is transformed as $\wh{\zeta} \to \wh{\zeta}' = \wh{\zeta} + 2 \pi \ell r^2$
under the shift $z \to z\,\e^{2 \pi \I}$. 
Then we can read off the monodromy of $\wh{\rho} = - 1/\wh{\omega}$ in such a way as
\begin{align}
\wh{\rho} \ \to \ 
\wh{\rho}' \ &= \ 
\frac{(1 + 2 \pi \ell rs) \wh{\rho} + 2 \pi \ell r^2}{- 2 \pi \ell s^2 \wh{\rho} + (1 - 2 \pi \ell rs)}
\, .
\end{align}
Thus we confirm that the new complex structure $\wh{\rho}$ is subject to the conjugate monodromy (\ref{conjugate-522}).
Applying $\tau = \I$ and $\wh{\rho}$ (\ref{rho3-522}) to the definition (\ref{field-config}), 
it turns out that the conjugate configuration is described as
\bsubeq \label{ConjugateConfig-522}
\begin{gather}
\d s^2 \ = \ 
\d s_{012345}^2 
+ H_{\ell} \big[ (\d \varrho)^2 + \varrho^2 (\d \vartheta)^2 \big]
+ \frac{\wh{\omega}_2}{|\wh{\omega}|^2} \big[ (\d y^8)^2 + (\d y^9)^2 \big]
\, , \\
\wh{B}_{89} 
\ = \ 
- \frac{\wh{\omega}_1}{|\wh{\omega}|^2}
\, , \ls
\e^{2 \wh{\phi}} 
\ = \ 
\frac{\wh{\omega}_2}{|\wh{\omega}|^2}
\, , \\
\wh{\omega}_1 \ = \
- \frac{\wh{\rho}_1}{|\wh{\rho}|^2}
\ = \ 
\frac{s}{r} - \frac{V_{\ell}}{r^2 K_{\ell}}
\, , \ls
\wh{\omega}_2 \ = \ 
\frac{\wh{\rho}_2}{|\wh{\rho}|^2}
\ = \ 
\frac{H_{\ell}}{r^2 K_{\ell}}
\, .
\end{gather}
\esubeq
This is the local expression of a defect $(r,-s)$ five-brane,
i.e., the composite of $r$ defect NS5-branes (\ref{dNS5-system}) and $-s$ exotic $5^2_2$-branes.
This is similar to the previous form (\ref{ConjugateConfig-dNS5}), 
while the roles of conjugate complex structures are different.
Indeed, the configuration (\ref{ConjugateConfig-522}) is generated via the T-duality transformations along the 8-th and 9-th direction of (\ref{ConjugateConfig-dNS5}), with relabeling $(p,q)$ to $(-s,r)$.
Simultaneously, the complex structures $(\tau, \wh{\rho})$ in (\ref{ConjugateConfig-dNS5}) are changed to $(\tau, \wh{\omega})$ in (\ref{ConjugateConfig-522}).

We again argue the mass of the composite system (\ref{ConjugateConfig-522}).
Parallel to the previous discussion in subsection \ref{cc-dNS5},
the dilaton is very small in the supergravity regime,
\begin{align}
g_{\text{st}}^2 \ &= \ 
\e^{2 \wh{\phi}} \ = \
\frac{\wh{\omega}_2}{|\wh{\omega}|^2}
\ \ll \ 1
\, .
\end{align}
This value also expresses the property of the radii $\wt{R}_{8,9}$ of the physical coordinates $y^{8,9}$ in (\ref{ConjugateConfig-522}),
\begin{align}
\Big( \frac{\wt{R}_8}{\ell_{\text{st}}} \Big)^2 
\ = \ 
\Big( \frac{\wt{R}_9}{\ell_{\text{st}}} \Big)^2 
\ = \ 
\frac{\wh{\omega}_2}{|\wh{\omega}|^2}
\ \ll \ 1
\, ,
\end{align}
in the appropriately large $\varrho$ region.
In this regime, the mass of the single defect NS5-brane and the single exotic $5^2_2$-brane are described as
\begin{align}
M_{\text{NS}} \ &= \ 
\frac{1}{g_{\text{st}}^2 \ell_{\text{st}}^6}
\, , \ls
M_{\text{E}} \ = \ 
\frac{1}{g_{\text{st}}^2 \ell_{\text{st}}^2 (\wt{R}_8 \wt{R}_9)^2}
\, , \ls
\frac{M_{\text{NS}}}{M_{\text{E}}}
\ = \ 
\frac{(\wt{R}_8 \wt{R}_9)^2}{\ell_{\text{st}}^4}
\ = \ 
\Big( \frac{\wh{\omega}_2}{|\wh{\omega}|^2} \Big)^2 
\ \ll \ 1
\, . 
\end{align}
Then it turns out that the mass of $-s$ exotic $5^2_2$-branes is dominant and the mass of $r$ defect NS5-branes is negligible.

\section{Summary and discussions}
\label{sect:summary}


In this paper, 
we studied the $SL(2,{\mathbb Z})_{\tau} \times SL(2,{\mathbb Z})_{\rho}$ monodromy structures of various defect five-branes.
We also investigated the conjugate configurations of them 
by virtue of the conjugate monodromy matrices and the corresponding complex structures.
Once we found the explicit forms of the complex structures which reproduce the conjugate monodromies, we immediately constructed the field configurations of the conjugate system.
In this process we constructed the metric, the B-field and the dilaton for the 
defect $(p,q)$ five-branes,
i.e., the composite of $p$ defect NS5-branes and $q$ exotic $5^2_2$-branes, in a concrete manner.
Since the configuration of the single defect five-brane is not globally well-defined,
the expression of the composite system would be quite helpful to find the globally well-defined formulation of defect five-branes, 
as in the case of seven-branes in type IIB theory.
If we find the global description for defect five-branes,
we will understand the importance of the exotic $5^2_2$-brane in a deeper level.
In this work we also obtained a new example of hyper-K\"{a}hler geometry (\ref{ConjugateConfig-dKK5}) as the conjugate system of the defect KK5-branes.
This is an ALG space, a generalization of an ALF space which asymptotically has a tri-holomorphic two-torus $T^{89}$ action \cite{Cherkis:2000cj, Cherkis:2001gm}.
Since this is originated from the Taub-NUT space via the smearing and the conjugating,
we can interpret this as the conjugated defect Taub-NUT space.


There are several discussions.
First, all of the conjugated configurations represent the composites of coincident defect five-branes.
In order to find a further general configuration where each defect five-brane is located at arbitrary point in the 67-plane,
we should introduce new parameters into the system.
In the case of multi-centered five-branes of codimension three,
we have already known the harmonic function,
\begin{align}
H \ = \ 1 + \frac{\ell_0}{\sqrt{2} |\vec{x}|}
\ \to \ 
1 + \sum_{k} \frac{\ell_0}{\sqrt{2} |\vec{x} - \vec{q}_k|}
\, ,  
\end{align}
where $\vec{q}_k$ is the position of the $k$-th five-brane in the 678-directions.
In the case of defect five-branes, however, it is difficult to recognize 
the sum of the harmonic functions for the single defect five-brane
as 
the one for multiple defect five-branes.
This is because the individual harmonic function involves the renormalization scale $\mu$.
Then we have to control many number of scale parameters in order that the harmonic function is well-defined in the whole region of the 67-plane.
This estimate is too naive to describe a number of separated defect five-branes.
In order to acquire the consistent form for such a configuration, 
we have to find the globally well-defined function 
as the modular invariant function for seven-branes in type IIB theory.
Nevertheless, the local descriptions of composite defect five-branes would be helpful to construct the globally well-defined function.

Apart from the above current difficulty, we can still study other topics.
(i) In the previous work \cite{Kimura:2013khz}, 
we tried to construct the GLSM for two defect five-branes.
Since we obtained the explicit configuration for them in the current work,
it can be possible to find the improved version of \cite{Kimura:2013khz}.
In particular, it would be quite interesting to construct string worldsheet theories for ALG spaces related to the previous work \cite{Kimura:2013fda}, by virtue of the T-duality transformation rules on the GLSM \cite{Kimura:2014bxa, Kimura:2014aja}.
(ii) In the system of the defect $(p,q)$ five-brane (\ref{ConjugateConfig-dNS5}), the 9-th direction is compactified and smeared.
Let us consider the string worldsheet instanton corrections along the 9-th direction.
From the viewpoint of the defect NS5-branes, the worldsheet instanton corrections to the 9-th direction can be interpreted as the KK momentum corrections \cite{Tong:2002rq, Harvey:2005ab, Okuyama:2005gx}.
On the other hand, from the viewpoint of the exotic $5^2_2$-branes, 
the worldsheet instanton corrections can be understood as the winding corrections to the configuration \cite{Kimura:2013zva}.
Then, how should we interpret the worldsheet instanton corrections to the defect $(p,q)$ five-brane?
This is quite a fascinating question.
(iii) There are various hyper-K\"{a}hler geometries.
If some of them coincide with the conjugate configurations discussed in this paper, 
we would be able to find a novel relation among various five-branes from the (non)geometrical viewpoint \cite{Kimura:2014upa, Kimura:2014bea}.

\section*{Acknowledgements}

The author would like to thank
Shin Sasaki 
and 
Masaya Yata
for valuable discussions and comments. 
This work is supported in part by the Iwanami-Fujukai Foundation.

\begin{appendix}

\section*{Appendix}

\section{T-duality}
\label{app:Buscher}

In this appendix we exhibit the T-duality transformations in two ways.
The first expression is the Buscher rule \cite{Buscher:1987sk}
of the T-duality transformation along the $n$-th direction,
\bsubeq \label{Buscher}
\begin{gather}
G'_{MN} \ = \ 
G_{MN} 
- \frac{G_{n M} G_{n N} - B_{n M} B_{n N}}{G_{nn}}
\, , \ls
G'_{n N} \ = \ 
\frac{B_{n N}}{G_{nn}}
\, , \ls
G'_{nn} \ = \ 
\frac{1}{G_{nn}}
\, , \\
B'_{MN} \ = \ 
B_{MN} 
+ \frac{2 G_{n [M} B_{N]n}}{G_{nn}}
\, , \ls
B'_{n N} \ = \ 
\frac{G_{n N}}{G_{nn}}
\, , \\
\phi' \ = \ 
\phi - \half \log (G_{nn})
\, .
\end{gather}
\esubeq
In the main part of this paper we frequently utilize this rule.
The explicit form is necessary for avoiding any sign ambiguities from involution of the B-field.

The second expression is a part of the monodromy transformations (\ref{MM-MM}).
The T-duality transformations along the 8-th and 9-th directions are represented in terms of $4 \times 4$ matrices $U_{8,9}$ in such a way as
\bsubeq \label{T-Umatrix}
\begin{align}
U_8 \ &= \ 
\left(
\begin{array}{cc}
\mathbbm{1} - T_8 & - T_8
\\
- T_8 & \mathbbm{1} - T_8
\end{array}
\right)
\, , \ls
T_8 \ \equiv \ 
\left(
\begin{array}{cc}
1 & 0 
\\
0 & 0
\end{array}
\right)
\, , \label{T8-M} \\
U_9 \ &= \ 
\left(
\begin{array}{cc}
\mathbbm{1} - T_9 & - T_9
\\
- T_9 & \mathbbm{1} - T_9
\end{array}
\right)
\, , \ls
T_9 \ \equiv \ 
\left(
\begin{array}{cc}
0 & 0 
\\
0 & 1
\end{array}
\right)
\, , \label{T9-M} \\
U_{89} \ &= \ 
\left(
\begin{array}{cc}
\mathbbm{1} - T_8 - T_9 & - T_8 - T_9
\\
- T_8 - T_9 & \mathbbm{1} - T_8 - T_9
\end{array}
\right)
\ = \ 
U_8 U_9 \ = \ 
U_9 U_8
\, . \label{T89-M}
\end{align}
\esubeq
In terms of these matrices, 
we can see the T-duality relations among the monodromy matrices
$\Omega^{\text{NS}}$, $\Omega^{\text{KK}}$ and $\Omega^{\text{E}}$,
\begin{align}
\Omega^{\text{KK}}
\ &= \ 
U_9^{\T} \, \Omega^{\text{NS}} \, U_9^{\vphantom{\T}}
\, , \ls
\Omega^{\text{E}}
\ = \ 
U_8^{\T} \, \Omega^{\text{KK}} \, U_8^{\vphantom{\T}}
\, , \ls
\Omega^{\text{E}}
\ = \ 
U_{89}^{\T} \, \Omega^{\text{NS}} \, U_{89}^{\vphantom{\T}}
\, .
\end{align}
The matrix description of the T-duality transformations can be also seen in \cite{Albertsson:2008gq, Kikuchi:2012za}, and so forth.

\section{Another configuration of defect KK5-brane}
\label{app:AK5}

In this appendix, 
we discuss another defect KK5-brane of different type\footnote{We note that the GLSM formulation is mentioned in appendix B of \cite{Kimura:2013fda}.}.
If we take T-duality along the 8-th direction of the defect NS5-brane (\ref{dNS5-system}), we obtain the following configuration, 
\bsubeq \label{adKK5-system}
\begin{gather}
\d s^2 \ = \ 
\d s_{012345}^2
+ H_{\ell} \, \big[ (\d \varrho)^2 + \varrho^2 (\d \vartheta)^2 \big]
+ H_{\ell} \, (\d x^9)^2 
+ \frac{1}{H_{\ell}} \big[ \d y^8 + V_{\ell} \, \d x^9 \big]^2 
\, , \\
B_{MN} \ = \ 0
\, , \ls 
\e^{2 \phi} \ = \ 1
\, .
\end{gather}
\esubeq
This is similar to the defect KK5-brane (\ref{dKK5-system}), 
while the structure of the two-torus $T^{89}$ is different.
In order to study the structure of the defect KK5-brane (\ref{adKK5-system}),
we analyze the matrix $\mathscr{M}^{\text{AK}}$ and the $O(2,2;{\mathbb Z})$ monodromy matrix
$\Omega^{\text{AK}}$ defined in (\ref{MM-MM}),
\begin{align}
\mathscr{M}^{\text{AK}} (\varrho, \vartheta)
\ &= \ 
\frac{1}{H_{\ell}}
\left(
\begin{array}{cccc}
1 & V_{\ell} & 0 & 0 
\\
V_{\ell} & K_{\ell} & 0 & 0 
\\
0 & 0 & K_{\ell} & - V_{\ell}
\\
0 & 0 & - V_{\ell} & 1
\end{array}
\right)
\, , \ls 
\Omega^{\text{AK}}
\ = \ 
\left(
\begin{array}{cccc}
1 & 2 \pi \ell & 0 & 0 
\\
0 & 1 & 0 & 0 
\\
0 & 0 & 1 & 0 
\\
0 & 0 & - 2 \pi \ell & 1
\end{array}
\right)
\, . 
\label{SO22-MM-adKK5}
\end{align}
The $O(2,2;{\mathbb Z})$ monodromy matrix $\Omega^{\text{AK}}$ is related to 
that of the defect NS5-brane $\Omega^{\text{NS}}$ (\ref{SO22-MM-dNS5})
and the exotic $5^2_2$-brane $\Omega^{\text{E}}$ (\ref{SO22-MM-522}) 
under the T-duality transformations (\ref{T-Umatrix}), 
\begin{align}
\Omega^{\text{AK}}
\ &= \ 
U_8^{\T} \, \Omega^{\text{NS}} \, U_8^{\vphantom{\T}}
\, , \ls
\Omega^{\text{E}}
\ = \ 
U_9^{\T} \, \Omega^{\text{AK}} \, U_9^{\vphantom{\T}}
\, . 
\end{align}

We can also discuss the $SL(2,{\mathbb Z})_{\tau} \times SL(2,{\mathbb Z})_{\rho}$ monodromy matrices
by virtue of the two-torus $T^{89}$ and two complex structures $\tau$ and $\rho$ defined by (\ref{field-config}).
Plugging (\ref{adKK5-system}) into (\ref{field-config}), we find
\begin{align}
\tau \ &= \ 
V_{\ell} + \I H_{\ell}
\ = \ 
\I h + \I \ell \log (\mu/z)
\, , \ls
\rho \ = \ \I
\, . 
\end{align}
Their monodromy transformations by the shift $z \to z \, \e^{2 \pi \I}$
are given as 
\bsubeq \label{adKK5-SL2SL2-monodromy}
\begin{alignat}{2}
\tau \ &\to \ 
\tau' \ = \ \tau + 2 \pi \ell 
\, , &\ls
\Omega_{\tau}^{\text{AK}}
\ &\equiv \ 
\left(
\begin{array}{cc}
1 & 2 \pi \ell \\
0 & 1
\end{array}
\right)
\, , \\
\rho \ &\to \ 
\rho' \ = \ \rho 
\, , &\ls
\Omega_{\rho}^{\text{AK}}
\ &\equiv \ 
\left(
\begin{array}{cc}
1 & 0 \\
0 & 1
\end{array}
\right)
\, .
\end{alignat}
\esubeq
This is similar to the monodromy matrices of the defect KK5-brane (\ref{dKK5-SL2SL2-monodromy}).
The slight difference of $\Omega_{\tau}^{\text{AK}}$ from $\Omega_{\tau}^{\text{KK}}$ originates in the difference of the involution of the KK-vector $\vec{V}$ into the off-diagonal part of the metric on the two-torus.

Let us now study the conjugate system by transformations of the monodromy matrices via (\ref{conjugate-U}),
\begin{align}
\wt{\Omega}_{\tau}^{\text{AK}}
\ &= \ 
\left(
\begin{array}{cc}
1 + 2 \pi \ell p'q' & 2 \pi \ell p'{}^2 
\\
- 2 \pi \ell q'{}^2 & 1 - 2 \pi \ell p'q'
\end{array}
\right)
\, , \ls
\wt{\Omega}_{\rho}^{\text{AK}}
\ = \ 
\left(
\begin{array}{cc}
1 & 0 \\
0 & 1
\end{array}
\right)
\, . \label{conjugate-adKK5}
\end{align}
Simultaneously, we also transform the complex structure $\tau$ in the following way,
\begin{align}
U_{\tau}^{-1} \ = \ 
\left(
\begin{array}{cc}
p' & -r' \\
- q' & s'
\end{array}
\right)
\, , \ls
\tau \ \to \ 
\wt{\tau} \ &\equiv \
\frac{p' \tau - r'}{-q' \tau + s'}
\, . \label{conjugate-tau-adKK5}
\end{align}
Since the original $\tau$ is transformed as $\tau \to \tau' = \tau + 2 \pi \ell$
under the shift $z \to z \, \e^{2 \pi \I}$,
then we can read off the monodromy of the new complex structure $\wt{\tau}$
in such a way as
\begin{align}
\wt{\tau} \ \to \ \wt{\tau}'
\ &= \
\frac{p' \tau' - r'}{-q' \tau' + s'}
\ = \
\frac{(1 + 2 \pi \ell p'q') \wt{\tau} + 2 \pi \ell p'{}^2}{- 2 \pi \ell q'{}^2 \wt{\tau} + (1 - 2 \pi \ell p'q')}
\, . 
\end{align}
This indicates that $\wt{\tau}$ is also subject to the conjugate monodromy (\ref{conjugate-adKK5}).
Then we find the generic form of the metric, the B-field and the dilaton for the composite configuration of $q'$ defect KK5-branes (\ref{dKK5-system}) and $p'$ defect KK5-branes (\ref{adKK5-system}),
\bsubeq \label{conjugate-BGphi-adKK5}
\begin{align}
\wt{G}_{88} 
\ &= \ 
\frac{s'{}^2 - 2 q' s' V_{\ell} + q'{}^2 K_{\ell}}{H_{\ell}}
\, , \\
\wt{G}_{89} \ = \ \wt{G}_{98}
\ &= \ 
- \frac{r's' - (p's' + r'q') V_{\ell} + p'q' K_{\ell}}{H_{\ell}}
\, , \\
\wt{G}_{99}
\ &= \ 
\frac{r'{}^2 - 2 p'r' V_{\ell} + p'{}^2 K_{\ell}}{H_{\ell}}
\, , \\
\wt{B}_{MN} \ &= \ 0
\, , \ls
\e^{2 \wt{\phi}} \ = \ 1
\, .
\end{align}
\esubeq
We note that the single defect KK5-brane (\ref{dKK5-system}) is realized by setting $(p',q',r',s') = (0,1,-1,0)$,
and the single defect KK5-brane of another type (\ref{adKK5-system}) is obtained by $(p',q',r',s') = (1,0,0,1)$.
However, the expression (\ref{conjugate-BGphi-adKK5}) is redundant for the case of generic $(q',p') \neq (0,0)$.
In order to find the reduced form for the generic $(q',p')$ configuration,
we rewrite the complex structure $\wt{\tau}$, 
\begin{align}
\wt{\tau} \ &= \ 
\frac{p' (V_{\ell} + \I H_{\ell}) - r'}{-q' (V_{\ell} + \I H_{\ell}) + s'}
\ = \ 
- \frac{p'}{q'} - \frac{1}{q' [(-s' + q' V_{\ell}) + \I q' H_{\ell}]}
\, . \label{tau2-adKK5}
\end{align}
Here we used $s'p' - q'r' = 1$ to eliminate $r'$.
Now the parameter $s'$ is arbitrary without any constraints from $(q',p')$.
Then we can set $s' = 0$ without loss of generality,
\begin{align}
\wh{\tau} \ &\equiv \ 
- \frac{p'}{q'} - \frac{1}{q'{}^2 (V_{\ell} + \I H_{\ell})}
\, . \label{tau3-adKK5}
\end{align}
For convenience,
we define $\wh{\zeta} \equiv -1/(\wh{\tau} + \frac{p'}{q'})$.
This is transformed as 
$\wh{\zeta} \to \wh{\zeta}' = \wh{\zeta} + 2 \pi \ell q'{}^2$
under the shift $z \to z \, \e^{2 \pi \I}$. 
Then we can read off the transformation of $\wh{\tau}$ in such a way as
\begin{align}
\wh{\tau} \ \to \ 
\wh{\tau}' \ &= \ 
\frac{(1 + 2 \pi \ell p'q') \wh{\tau} + 2 \pi \ell p'{}^2}{- 2 \pi \ell q'{}^2 \wh{\tau} + (1 - 2 \pi \ell p'q')}
\, .
\end{align}
Then we find that $\wh{\tau}$ is the conjugate complex structure of the conjugate monodromy (\ref{conjugate-adKK5}).
Substituting $\rho = \I$ and $\wh{\tau}$ (\ref{tau3-adKK5}) into the definition (\ref{field-config}),
we also find the local expression of the conjugate system,
\bsubeq \label{ConjugateConfig-adKK5}
\begin{gather}
\d s^2 \ = \ 
\d s_{012345}^2 
+ H_{\ell} \big[ (\d \varrho)^2 + \varrho^2 (\d \vartheta)^2 \big]
+ \frac{1}{\wh{\tau}_2} \big[ \d y^8 + \wh{\tau}_1 \, \d x^9 \big]^2
+ \wh{\tau}_2 \, (\d x^9)^2
\, , \\
\wh{B}_{MN} \ = \ 0
\, , \ls
\e^{2 \wh{\phi}} \ = \ 1
\, , \\
\wh{\tau}_1 \ = \ 
- \frac{p'}{q'} - \frac{V_{\ell}}{q'{}^2 K_{\ell}}
\, , \ls
\wh{\tau}_2 \ = \ 
\frac{H_{\ell}}{q'{}^2 K_{\ell}}
\, .
\end{gather}
\esubeq
Close to the situation of (\ref{ConjugateConfig-dKK5}),
the transverse space of the 6789-directions is also a hyper-K\"{a}hler geometry of ALG type \cite{Cherkis:2000cj, Cherkis:2001gm}.
This system is also found via the T-duality transformation along the 8-th direction of the conjugate system of the defect NS5-brane (\ref{ConjugateConfig-dNS5}),
where we also relabel $(p,q)$ in (\ref{ConjugateConfig-dNS5}) to $(p',q')$ in (\ref{ConjugateConfig-adKK5}).

\end{appendix}

}

\begin{thebibliography}{99}
\setlength{\itemsep}{.9mm}

\bibitem{Strominger:1990et}
  A.~Strominger,
  {\sl Heterotic solitons},
  Nucl.\ Phys.\ B {\bf 343} (1990) 167
   [Erratum-ibid.\ B {\bf 353} (1991) 565].

\bibitem{Callan:1991dj}
  C.~G.~Callan, Jr., J.~A.~Harvey and A.~Strominger,
  {\sl Worldsheet approach to heterotic instantons and solitons},
  Nucl.\ Phys.\ B {\bf 359} (1991) 611.

\bibitem{Hanany:1996ie}
  A.~Hanany and E.~Witten,
  {\sl Type IIB superstrings, BPS monopoles, and three-dimensional gauge dynamics},
  Nucl.\ Phys.\ B {\bf 492} (1997) 152
  [hep-th/9611230].

\bibitem{Giveon:1998sr}
  A.~Giveon and D.~Kutasov,
  {\sl Brane dynamics and gauge theory},
  Rev.\ Mod.\ Phys.\  {\bf 71} (1999) 983
  [hep-th/9802067].

\bibitem{Gaiotto:2009we}
  D.~Gaiotto,
  {\sl $\N=2$ dualities},
  JHEP {\bf 1208} (2012) 034
  [arXiv:0904.2715 [hep-th]].

\bibitem{Sorkin:1983ns}
  R.~D.~Sorkin,
  {\sl Kaluza-Klein monopole},
  Phys.\ Rev.\ Lett.\  {\bf 51} (1983) 87.

\bibitem{deBoer:2010ud}
  J.~de Boer and M.~Shigemori,
  {\sl Exotic branes and non-geometric backgrounds},
  Phys.\ Rev.\ Lett.\  {\bf 104} (2010) 251603
  [arXiv:1004.2521 [hep-th]].

\bibitem{deBoer:2012ma}
  J.~de Boer and M.~Shigemori,
  {\sl Exotic branes in string theory},
  Phys.\ Rept.\  {\bf 532} (2013) 65
  [arXiv:1209.6056 [hep-th]].

\bibitem{Gregory:1997te}
  R.~Gregory, J.~A.~Harvey and G.~W.~Moore,
  {\sl Unwinding strings and T-duality of Kaluza-Klein and H-monopoles},
  Adv.\ Theor.\ Math.\ Phys.\  {\bf 1} (1997) 283
  [hep-th/9708086].

\bibitem{Hori:2002cd}
  K.~Hori and A.~Kapustin,
  {\sl Worldsheet descriptions of wrapped NS five-branes},
  JHEP {\bf 0211} (2002) 038
  [hep-th/0203147].

\bibitem{Tong:2002rq}
  D.~Tong,
  {\sl NS5-branes, T-duality and worldsheet instantons},
  JHEP {\bf 0207} (2002) 013
  [hep-th/0204186].

\bibitem{Harvey:2005ab}
  J.~A.~Harvey and S.~Jensen,
  {\sl Worldsheet instanton corrections to the Kaluza-Klein monopole},
  JHEP {\bf 0510} (2005) 028
  [hep-th/0507204].

\bibitem{Okuyama:2005gx}
  K.~Okuyama,
  {\sl Linear sigma models of H and KK monopoles},
  JHEP {\bf 0508} (2005) 089
  [hep-th/0508097].

\bibitem{Bergshoeff:2011se}
  E.~A.~Bergshoeff, T.~Ort\'{\i}n and F.~Riccioni,
  {\sl Defect branes},
  Nucl.\ Phys.\ B {\bf 856} (2012) 210
  [arXiv:1109.4484 [hep-th]].

\bibitem{Kimura:2013fda}
  T.~Kimura and S.~Sasaki,
  {\sl Gauged linear sigma model for exotic five-brane},
  Nucl.\ Phys.\ B {\bf 876} (2013) 493
  [arXiv:1304.4061 [hep-th]].

\bibitem{Kimura:2013zva}
  T.~Kimura and S.~Sasaki,
  {\sl Worldsheet instanton corrections to $5^2_2$-brane geometry},
  JHEP {\bf 1308} (2013) 126
  [arXiv:1305.4439 [hep-th]].

\bibitem{Chatzistavrakidis:2013jqa}
  A.~Chatzistavrakidis, F.~F.~Gautason, G.~Moutsopoulos and M.~Zagermann,
  {\sl Effective actions of non-geometric fivebranes},
  Phys.\ Rev.\ D {\bf 89} (2014) 066004
  [arXiv:1309.2653 [hep-th]].

\bibitem{Kimura:2014upa}
  T.~Kimura, S.~Sasaki and M.~Yata,
  {\sl World-volume effective actions of exotic five-branes},
  JHEP {\bf 1407} (2014) 127
  [arXiv:1404.5442 [hep-th]].

\bibitem{Bergshoeff:1997gy}
  E.~Bergshoeff, B.~Janssen and T.~Ortin,
  {\sl Kaluza-Klein monopoles and gauged sigma models},
  Phys.\ Lett.\ B {\bf 410} (1997) 131
  [hep-th/9706117].

\bibitem{Greene:1989ya} 
  B.~R.~Greene, A.~D.~Shapere, C.~Vafa and S.~T.~Yau,
  {\sl Stringy cosmic strings and noncompact Calabi-Yau manifolds},
  Nucl.\ Phys.\ B {\bf 337} (1990) 1.

\bibitem{DeWolfe:1998eu}
  O.~DeWolfe, T.~Hauer, A.~Iqbal and B.~Zwiebach,
  {\sl Uncovering the symmetries on $[p,q]$ seven-branes: beyond the Kodaira classification},
  Adv.\ Theor.\ Math.\ Phys.\  {\bf 3} (1999) 1785
  [hep-th/9812028].

\bibitem{DeWolfe:1998pr}
  O.~DeWolfe, T.~Hauer, A.~Iqbal and B.~Zwiebach,
  {\sl Uncovering infinite symmetries on $[p, q]$ 7-branes: Kac-Moody algebras and beyond},
  Adv.\ Theor.\ Math.\ Phys.\  {\bf 3} (1999) 1835
  [hep-th/9812209].

\bibitem{Bergshoeff:2006jj}
  E.~A.~Bergshoeff, J.~Hartong, T.~Ort\'{\i}n and D.~Roest,
  {\sl Seven-branes and supersymmetry},
  JHEP {\bf 0702} (2007) 003
  [hep-th/0612072].

\bibitem{Hellerman:2002ax}
  S.~Hellerman, J.~McGreevy and B.~Williams,
  {\sl Geometric constructions of nongeometric string theories},
  JHEP {\bf 0401} (2004) 024
  [hep-th/0208174].

\bibitem{Kikuchi:2012za}
  T.~Kikuchi, T.~Okada and Y.~Sakatani,
  {\sl Rotating string in doubled geometry with generalized isometries},
  Phys.\ Rev.\ D {\bf 86} (2012) 046001
  [arXiv:1205.5549 [hep-th]].

\bibitem{Cherkis:2000cj} 
  S.~A.~Cherkis and A.~Kapustin,
  {\sl Nahm transform for periodic monopoles and $\N=2$ super Yang-Mills theory},
  Commun.\ Math.\ Phys.\  {\bf 218} (2001) 333
  [hep-th/0006050].

\bibitem{Cherkis:2001gm}
  S.~A.~Cherkis and A.~Kapustin,
  {\sl Hyperk\"{a}hler metrics from periodic monopoles},
  Phys.\ Rev.\ D {\bf 65} (2002) 084015
  [hep-th/0109141].

\bibitem{Jensen:2011jna}
  S.~Jensen,
  {\sl The KK-monopole/NS5-brane in doubled geometry},
  JHEP {\bf 1107} (2011) 088
  [arXiv:1106.1174 [hep-th]].

\bibitem{Berman:2014jsa}
  D.~S.~Berman and F.~J.~Rudolph,
  {\sl Branes are waves and monopoles},
  arXiv:1409.6314 [hep-th].

\bibitem{Andriot:2014uda}
  D.~Andriot and A.~Betz,
  {\sl NS-branes, source corrected Bianchi identities, and more on backgrounds with non-geometric fluxes},
  JHEP {\bf 1407} (2014) 059
  [arXiv:1402.5972 [hep-th]].

\bibitem{Maharana:1992my}
  J.~Maharana and J.~H.~Schwarz,
  {\sl Noncompact symmetries in string theory},
  Nucl.\ Phys.\ B {\bf 390} (1993) 3
  [hep-th/9207016].

\bibitem{Hull:2004in}
  C.~M.~Hull,
  {\sl A geometry for non-geometric string backgrounds},
  JHEP {\bf 0510} (2005) 065
  [hep-th/0406102].

\bibitem{Lawrence:2006ma}
  A.~Lawrence, M.~B.~Schulz and B.~Wecht,
  {\sl D-branes in nongeometric backgrounds},
  JHEP {\bf 0607} (2006) 038
  [hep-th/0602025].

\bibitem{Hull:2006qs}
  C.~M.~Hull,
  {\sl Global aspects of T-duality, gauged sigma models and T-folds},
  JHEP {\bf 0710} (2007) 057
  [hep-th/0604178].

\bibitem{Dall'Agata:2007sr}
  G.~Dall'Agata, N.~Prezas, H.~Samtleben and M.~Trigiante,
  {\sl Gauged supergravities from twisted doubled tori and non-geometric string backgrounds},
  Nucl.\ Phys.\ B {\bf 799} (2008) 80
  [arXiv:0712.1026 [hep-th]].

\bibitem{Kimura:2013khz}
  T.~Kimura and S.~Sasaki,
  {\sl Worldsheet description of exotic five-brane with two gauged isometries},
  JHEP {\bf 1403} (2014) 128
  [arXiv:1310.6163 [hep-th]].

\bibitem{Kimura:2014bxa}
  T.~Kimura and M.~Yata,
  {\sl Gauged linear sigma model with F-term for A-type ALE space},
  PTEP {\bf 2014} (2014) 7,  073B01
  [arXiv:1402.5580 [hep-th]].

\bibitem{Kimura:2014aja}
  T.~Kimura and M.~Yata,
  {\sl T-duality transformation of gauged linear sigma model with F-term},
  Nucl.\ Phys.\ B {\bf 887} (2014) 136
  [arXiv:1406.0087 [hep-th]].

\bibitem{Kimura:2014bea}
  T.~Kimura, S.~Sasaki and M.~Yata,
  {\sl Hyper-K\"{a}hler with torsion, T-duality, and defect $(p,q)$ five-branes},
  arXiv:1411.3457 [hep-th].

\bibitem{Buscher:1987sk}
  T.~H.~Buscher,
  {\sl A symmetry of the string background field equations},
  Phys.\ Lett.\ B {\bf 194} (1987) 59.


\bibitem{Albertsson:2008gq}
  C.~Albertsson, T.~Kimura and R.~A.~Reid-Edwards,
 {\sl D-branes and doubled geometry},
  JHEP {\bf 0904} (2009) 113
  [arXiv:0806.1783 [hep-th]].

\end{thebibliography}
\end{document}